\RequirePackage{fix-cm}
\documentclass[twocolumn,epj]{svjour}  
\smartqed  
\RequirePackage{graphicx}
\journalname{Eur. Phys. J. C}

\usepackage{xcolor}
\usepackage{color}
\usepackage{epsf}
\usepackage{soul}
\usepackage{etex}
\usepackage{morefloats}

\colorlet{darkgreen}{green!50!black}
\colorlet{brightyellow}{yellow!75!red}
\colorlet{orange}{red!50!yellow}
\colorlet{darkblue}{blue!60!black}
\colorlet{darkred}{red!80!black}

\newcommand{\cd}{\makebox[0.08cm]{$\cdot$}}

\begin{document}

\title{Abnormal states with unequal constituent masses}

\author{V.A. Karmanov\inst{1}}
%
%
\institute{Lebedev Physical Institute, Leninsky Prospekt 53, 119991
Moscow, Russia}

\date{Received: date / Accepted: date}

\abstract{The Bethe-Salpeter equation for system of two oppositely charged particles 
 not only  reproduces the Coulomb spectrum, but, for enough large coupling constant ${\cal C}>\frac{\pi}{4}$, predicts additional levels  
 not  predicted by the Schr\"odinger equation.  These relativistic states (called abnormal), in contrast to the normal ones, are dominated, for
 more than 90-99 percent, by Fock states involving the exchange particles - the photons, whereas contribution of two massive charged particles themselves is rather small (1-10 \%).
Since the carrier of a large (positive) charge is a heavy ion, and the negative charge is provided by electron,  the masses of 
two constituents are very different. 
It is shown that in a system with so different masses 
the abnormal states still exist. Moreover, the effect of unequal masses is attractive. 
The balance between photons and charged constituents is  weakly sensitive to the mass ratio, 
so the photons still predominate.}

\maketitle


\section{Introduction}\label{intr}
The Bethe-Salpeter (BS) equation \cite{bs} for a hydrogen atom reproduces the Cou\-lomb levels with the relativistic corrections.
However, spectrum of the BS equation for electron in strong Coulomb field contains more than the usual Coulomb states.
In their seminal studies \cite{Wick,Cutkosky} of properties of the  BS equation  for two particles interacting via scalar massless exchange (reduced, in the non-relativistic case, to the Coulomb potential),  
Wick and Cutkosky have found that,
in addition to the Coulomb spectrum, there exist   levels which have no   non-relativistic counterparts. They appear if the coupling constant ${\cal C}$  is rather large: \mbox{${\cal C}>\frac{\pi}{4}$} (in the units when for the electron charge $e^2=\frac{1}{137}$). However, the energies of these levels are small (they are highly excited in full spectrum). These states, not predicted by the Schr\"odinger equation,  were called "abnormal". After their theoretical discovery, there was discussion devoted to their existence in
nature (see for review \cite{Nakanishi_prog}). Some researchers considered them as manifestation of an unknown defect of the BS equation. Other ones argued that since these levels are predicted by the BS equation on the same ground as the Coulomb ones, they cannot be ignored and must exist in nature. In their opinion (which we share), this is inconsistent to believe in one series of levels (the Coulomb one) and to deny the physical meaning of other series (overlapping with the Coulomb one), predicted by the same equation with the same interaction. The nature of these states was clarified in the recent paper \cite{CKS21}. 
It was found that the abnormal states, in contrast to the normal ones,  are dominated by the exchange particles - the photons  (which are scalar in the Wick-Cutkosky model). That's why they are  beyond the realm of the non-relativistic Schr\"odinger equation. 

This puts on the agenda the experimental detection of the abnormal states. Therefore the theoretical predictions should be made in situation as close as possible to an experimental one. The works \cite{Wick,Cutkosky,CKS21} were concentrated on the principal problems and therefore
studied the systems of the charged particles  with equal masses (though, the unequal masses equation was also presented in \cite{Cutkosky}).
The systems with two large opposite charges and equal masses are formed by nucleus-antinucleus  with a heavy enough antinucleus, to obtain a large coupling constant ${\cal C}$ as the product of two charges. At the present, these heavy antinuclei  can be hardly created and used in experiments. The system with large charge which can be studied in laboratory is a heavy ion and electron. This means that the constituents have very different masses.
Therefore, the systems containing light and heavy charged particles are of particular interest.

The equation for the weight function determining the BS amplitude was derived by Cutkosky \cite{Cutkosky} both for the cases of equal and unequal masses.
However,  in contrast to \cite{Cutkosky}, some  very  important "subtlety"  will  be taken into account  in our studying the spectrum  in the unequal masses case: $m_1\neq m_2$. Namely, we will show that
the coupling constant $\lambda$, entering in the Cutkosky equation, is related to the Coulomb constant ${\cal C}$ by a mass-dependent factor (see Eq. (\ref{lamal}) below) and, for fixed 
${\cal C}$,  for large values of the ratio
\begin{equation}\label{mratio}
r=\frac{m_1}{m_2},
\end{equation} 
the constant $\lambda$ decreases like $\sim\frac{1}{r}$. In Ref. \cite{Cutkosky} $\lambda$ was not related to the Coulomb constant ${\cal C}$ and assumed independent of the constituent masses. This was a different, mathematically correct, but unrealistic statement of problem.
 We imply that the particle 2 is electron, so the mass $m_2$ is fixed, whereas the ion mass $m_1$ is large relative to $m_2$.
It varies from ion to ion. For the transuranic elements, which are the only ones that can provide a sufficiently large charge, the ratio $r> 10^6$. In this case, the limit $r\to \infty$ is a very good approximation to the realistic situation. In principle, the decrease of $\lambda$ with increase of $r$ might eliminate the abnormal solutions. We will see that this does not happen.

In the present work, we will check the existence of the abnormal states in the system of unequal masses and study the dependence
of the level energies and the Fock sector content of the state vector on the  ratio $r$.
We will show that in spite of decrease of $\lambda$ vs. $r$,  the heavy ion and electron are still forming the abnormal states and the effect of unequal masses in comparison to the equal ones, due to other mass-dependent terms in the equation, leads even to an increase of the binding energy up to  a factor of two. This could make it easier to detect the abnormal states. 

Following Wick and Cutkosky, we don't include spin effects. Therefore the "photons" filling the abnormal states are scalar in the Wick-Cutkosky model. Inclusion of spin could precise the predictions, but it hardly removes the abnormal states since it
 does not eliminate strong Coulomb attraction and relativistic retardation of the numerous exchanged photons (i.e., finite flying time needed for exchange) 
 due to which they fill in the intermediate states and dominate the Fock sectors.
 
 In Sec. \ref{m1m2} we give a brief summary of the results \mbox{\cite{Wick,Cutkosky}} concerning the integral representation of the BS amplitude and the equation in the unequal masses case. In Sec. \ref{be}  dependence of the binding energy on the mass ratio $r$
 is studied. In Sec. \ref{N_2} the relation between the BS amplitude and the light-front two-body wave function is extended to the case of unequal masses.  The two-body constituent contribution to full normalization integral is calculated in Sec. \ref{sect_N2}. The corresponding normalization factor is found in Sec. \ref{norm}. Numerical results are given in Sec. \ref{num}. Concluding remarks are presented in 
 Sec. \ref{concl}.  The Appendices \ref{app1} and \ref{app2} contain some technical details -- the  the mentioned above $r$-dependent relation between the product of charges ${\cal C}$ used in the present paper and the coupling constant $\lambda$ used in Refs. \cite{Wick,Cutkosky}, as well as the relation between the four-vector scalar products and the light-front variables $\vec{R}_{\perp},x$.

\section{BS amplitude in the unequal-masses case}\label{m1m2}

Like in Ref. \cite{CKS21}, our consideration is based on the integral representation of the BS amplitude, found in \cite{Wick,Cutkosky} also for the unequal masses case, and on the equation for the weight function, determining this representation. In the present Section we give summary of these results.

In the unequal masses case $m_1\neq m_2$, for a general kernel $K$, the   equation for the BS amplitude $\Phi_{un}$ in Minkowski space  has the form:
\begin{eqnarray}\label{BSeq1}
&&(k_1^2-m_1^2)(k_2^2-m_2^2)\Phi_{un}(k_1,k_2;p)=
\nonumber\\
&-&i\int K(k_1,k'_1;p)\Phi_{un}(k'_1,k'_2;p)\delta^{(4)}(k'_1+k'_2-p)\frac{d^4k'_1}{(2\pi)^4}.
\nonumber\\
&&
\end{eqnarray}
We use the subscript "un" (abbreviation from "unequal")  to emphasize that this amplitude satisfies the equation with unequal masses.
Corresponding amplitude in the equal masses case will be denoted $\Phi_{eq}$.

Following Cutkosky \cite{Cutkosky},  we introduce the relative momentum $k$ as:
\begin{equation}\label{k}
k=\mu_2 k_1-\mu_1 k_2, 
\end{equation}
 where
 $$
 \mu_{1,2}=\frac{m_{1,2}}{m_1+m_2},\quad \mu_1+\mu_2=1.
$$
The particle momenta $k_{1,2}$ are expressed via $k$ and $p$:
\begin{equation}\label{k12}
k_1=\mu_1 p+k,\quad k_2=\mu_2 p-k,\quad k_1+k_2=p.
\end{equation}
We assume that the kernel $K(k_1,k'_1;p)$ corresponds to exchange by a massless particle.
Then, in the notations adopted in \cite{Cutkosky},  the equation (\ref{BSeq1}) is rewritten as:\footnote{We restore in r.h.-side of Eq. (\ref{BSeq2}) the factor $m^2_{12}$ which is  set to 1 in \cite{Cutkosky}.} 
\begin{eqnarray}\label{BSeq2}
&&[(\mu_1p+k)^2-m_1^2][(\mu_2p-k)^2-m_2^2]\Phi_{un}(k,p)
\nonumber\\
&=&\frac{i\lambda m^2_{12}}{\pi^2}\int\frac{\Phi_{un}(k';p)d^4k'}{(k-k')^2+i\epsilon},
\end{eqnarray}
where 
$$
m_{12}=\frac{1}{2}(m_1+m_2)
$$
and $p^2=M^2$, $M$ is the total mass of the bound system, determined by the equation (\ref{BSeq2}), $M=m_1+m_2-B$, $B$ is the module of the binding energy: $B=|E_b|$.
 
We deal with two particles with the charges $e_1$ and $e_2$ so that their interaction is determined by the product 
$$
{\cal C}=e_1 e_2.
$$ 
As mentioned above, we imply the units in which for electron $e^2\approx \frac{1}{137}$.
This Coulomb constant ${\cal C}$ will be used as input.
 Whereas, the BS equation (\ref{BSeq2}) contains the constant $\lambda$. 
The latter is not product of charges but, as we will see, differs from it by a factor. In Appendix  \ref{app1} we express $\lambda$ via ${\cal C}$. The relation (\ref{lamal0}) between these two constants  can be represented in the form
\begin{equation}\label{lamal}
\lambda=(1-\Delta^2)\frac{\cal C}{\pi}=\frac{4r}{\pi(r+1)^2}{\cal C},
\end{equation}
where
\begin{equation}\label{Delta}
\Delta=\frac{m_1-m_2}{m_1+m_2}=\frac{r-1}{r+1}
\end{equation}
and $r$ is defined in Eq. (\ref{mratio}).
That is\footnote{In the unequal masses case Cutkosky \cite{Cutkosky} takes: $m_1=1+\Delta$, $m_2=1-\Delta$ that implies $m_1+m_2=2$. 
We introduce the arbitrary masses $m_1,m_2$ explicitly.}
\begin{equation}\label{m12}
m_1=m_{12}(1+\Delta),\; m_2=m_{12}(1-\Delta).
\end{equation}
From Eq. (\ref{lamal}) it follows that for fixed ${\cal C}$ the value of $\lambda$ vs. $r>1$ decreases.

Comparing Eqs. (\ref{BSeq1}) and (\ref{BSeq2}), we find the kernel  in Eq. (\ref{BSeq1}):
\begin{eqnarray}\label{kern0}
K(k_1,k'_1;p)=K(k-k')&=&-\frac{16\pi^2\lambda m^2_{12}}{(k-k')^2+i\epsilon}
\\
&=&-\frac{ 16\pi{\cal C} m^2_{12}(1-\Delta^2)}{(k-k')^2+i\epsilon}.
\nonumber
\end{eqnarray}
Note, appearing due to the relation (\ref{lamal}), the factor $(1-\Delta^2)$  which will be cancelled in the equation for the weight function $g$.
This factor does not appear in the corresponding equation in Ref. \cite{Cutkosky}, 
 containing the constant $\lambda$. It drastically changes the spectrum when $r\to\infty$ ($\Delta\to 1$).

The BS amplitude  $\Phi=\Phi(k,p)$, both in equal and unequal masses cases, is characterized by the principal quantum number $n$. For the S-wave and $n=1$ it
has the following representation in the form of integral over the variable $z$ (Eqs. (12), (29) from \cite{Cutkosky}, taken for $n=1$,  and for the angular momentum $l=0$, rewritten in the Minkowski space and in different notations):
\begin{eqnarray}\label{Phi1}
&&\Phi_{un}(k,p)=-i m_{12}^3\int_{-1}^1dz \;g_{un}(z,\Delta)\times
\nonumber\\
&&\hspace{-0.3cm}\frac{1}{[m^2_{12}(1-\eta^2_{un})(1+2z\Delta+\Delta^2)-k^2-kp(z+\Delta)-i\epsilon]^3},
\nonumber\\
&&
\end{eqnarray}
where
$$
\eta^2_{un}=\frac{M^2}{4m^2_{12}},
$$
$M$ is the total mass of the system defined after Eq. (\ref{BSeq2}).
With the factor $m_{12}^3$ in (\ref{Phi1}) the function $g_{un}(z,\Delta)$ is dimensionless. 

With the relation (\ref{lamal}) between the coupling constants,
the BS equation (\ref{BSeq2})  is reduced to the following integral equation for the  auxiliary weight function $g_{un}(z,\Delta)$,  determining via the integral (\ref{Phi1}) 
the BS amplitude (see Eq. (30) from \cite{Cutkosky}):
\begin{equation}\label{eqgz2}
g_{un}(z,\Delta)=\frac{{\cal C}(1-\Delta^2)}{2\pi n}\int_{-1}^1 R^n(z,\xi)\frac{g_{un}(\xi,\Delta)}{Q(\xi,\Delta)}d\xi,
\end{equation}
where 
\begin{equation}\label{R}
R(z,\xi)=
\left\{
\begin{array}{ll}
\frac{1-z}{1-\xi},& \mbox{if $\xi<z$}
\\
\frac{1+z}{1+\xi},& \mbox{if $\xi>z$}
\end{array}
\right.
\end{equation}
and
\begin{equation}\label{Qt}
Q(\xi,\Delta)=(1+2\xi\Delta+\Delta^2)(1-\eta_{un}^2)+\eta_{un}^2(\xi+\Delta)^2.
\end{equation}
The equation (\ref{eqgz2}) is valid for the case of arbitrary principal quantum number $n$. The generalization of the integral representation (\ref{Phi1}) for this case  $n>1$ can be found in Refs. \cite{Cutkosky,CKS21}. 

The integral equation (\ref{eqgz2}) is equivalent to the following differential (with respect to $z$) equation (Eq. (31) from \cite{Cutkosky}):
\begin{eqnarray}\label{equn}
g''_{un}(z,\Delta)&+&\left[2(n-1)zg'_{un}(z,\Delta)-n(n-1)g_{un}(z,\Delta)\right.
\nonumber\\
&+&\left.\frac{{\cal C}(1-\Delta^2)}{\pi Q(z,\Delta)}\right]\frac{g_{un}(z,\Delta)}{(1-z^2)}=0,
\end{eqnarray}
with the boundary conditions $g_{un}(z=\pm 1,\Delta)=0$.

For the equal masses $m_1=m_2$ ($\Delta=0$) Eq. (\ref{Phi1}) turns into Eq. (6) from  \cite{CKS21} (taken for $n=1$ and without the normalization factor):
\begin{eqnarray}\label{Phi}
\Phi_{eq}(k,p)=\int_{-1}^1 \frac{-im_2^3g_{eq}(z)dz }
{[m_2^2(1-\eta^2_{eq}) -k^2-p\cd k\,z-i\epsilon]^{3}}     
\end{eqnarray}
with $\eta_{eq}^2=\frac{M^2}{4m_2^2}$.   As noted above, by the subscript "eq" we mark the quantities related to the case of equal masses.
Then the integral equation (\ref{eqgz2}) turns into 
\begin{equation}\label{eqg2b}
g_{eq}\left(z\right)=\frac{{\cal C}}{2\pi n}
\int_{-1}^1\frac{R^n(z,\xi) g_{eq}\left(\xi\right)}{\left[1-\eta_{eq}^2(1-\xi^2)\right]}d\xi,
\end{equation}
and the corresponding  differential one differs from (\ref{equn}) by the last term:
\begin{eqnarray}\label{eqeq}
g''_{eq}(z)&+&\left[2(n-1)zg'_{eq}(z)-n(n-1)g_{eq}(z)\phantom{\frac{\cal C}{\pi}}\right.
\nonumber\\
&+&\left.\frac{\cal C}{\pi [1-\eta_{eq}^2(1-z^2)]}\right]\frac{g_{eq}(z)}{(1-z^2)}=0.
\end{eqnarray}
The factor $(1-\Delta^2)$ at the front of the constant ${\cal C}$ in Eqs. (\ref{eqgz2}), (\ref{equn}) disappears in Eqs. (\ref{eqg2b}), (\ref{eqeq}).

The principal quantum number $n$ enters in Eqs. (\ref{eqgz2}), (\ref{equn}), (\ref{eqg2b}) and (\ref{eqeq}) as a parameter. For fixed $n$ all these 
homogeneous equations create another infinite series of levels labeled by $\kappa=0,1,2,3,\ldots$.  This is the origin of the abnormal solutions. The value $\kappa=0$ corresponds to the normal state, the
non-zero $\kappa$'s   correspond to the abnormal ones.  It was shown \cite{cia} that the states with odd $\kappa$ don't contribute to the S-matrix and
are not observable. Therefore we considered in Ref. \cite{CKS21} and  will consider in the present article the abnormal states with even 
$\kappa$ only.

Cutkosky has shown \cite{Cutkosky} that the unequal mass equation  (\ref{eqgz2}) can be transformed to the form of the equal mass one (\ref{eqg2b}).
This sophisticated  transformation is achi\-eved by the following replacement of variable and the function:\footnote{The replacement of variable  given in \cite{Cutkosky} below Eq. (31) contains a misprint. It differs from the correct one - Eqs.   (\ref{zzbar}), (\ref{gtilde1})  of the present paper - by the sign at the front of $\Delta$. 
In spite of that, the result   - Eq. (33) in \cite{Cutkosky} (Eq. (\ref{eqg2b}) in the present article) - is correct.}
\begin{eqnarray}
z&=&\frac{(\bar{z}-\Delta)}{(1-\Delta \bar{z})} \leftrightarrow \bar{z}=\frac{(z+\Delta)}{(1+\Delta z)},
\label{zzbar}
\\
g_{un}(z,\Delta)&=&\frac{g_{eq}(\bar{z})}{(1-\Delta\bar{z})^n}=\left(\frac{1+\Delta z}{1-\Delta^2}\right)^ng_{eq}\left(\frac{z+\Delta}{1+\Delta z}\right).
\nonumber\\
&&
\label{gtilde1}
\end{eqnarray}
After this replacement, the equation (\ref{eqgz2}) for $g_{un}(z,\Delta)$ turns into Eq. (\ref{eqg2b}) for $g_{eq}(z)$ with
\begin{equation}\label{etaeq}
\eta_{eq}^2=\frac{\eta_{un}^2-\Delta^2}{1-\Delta^2}.
\end{equation}
That is, in order to find the solution of the unequal masses equation (\ref{eqgz2}),
it is enough to solve the equal masses equation (\ref{eqg2b}).
Then the eigenvalues $\eta_{un}^2$, which enter in (\ref{eqgz2}) via $Q(\xi,\Delta)$, Eq. (\ref{Qt}), are expressed through $\eta_{eq}^2$, found as eigenvalues of Eq. (\ref{eqg2b}),  by the relation:
\begin{equation}\label{eta2un}
\eta_{un}^2=\Delta^2+(1-\Delta^2)\eta_{eq}^2
\end{equation}
and the solution $g_{un}(z,\Delta)$ of Eqs. (\ref{eqgz2}), (\ref{equn})  is expressed through the solution  $g_{eq}(z)$ of Eq. (\ref{eqg2b}) by the relation 
(\ref{gtilde1}).

We emphasize that the equal masses equations (\ref{eqg2b}) and (\ref{eqeq}), where $\eta^2_{eq}$ is considered as eigenvalue, don't contain any masses at all.  The non-equal masses equations (\ref{eqgz2}) and (\ref{equn}) contain the masses only in the form of the ratio $\Delta$. The particle masses 
(not their ratios) appear 
in the relations   between the eigenvalues $\eta_{eq}^2, \eta_{un}^2$ and the corresponding binding energies.

Note that  numerically solving the unequal masses differential equation (\ref{equn}) is not more complicated and is  as fast as solving the equal masses one (\ref{eqeq}).
There is no   practical advantage of Eq. (\ref{eqeq}) over (\ref{equn}) (or Eq. (\ref{eqg2b}) over (\ref{eqgz2})) in this respect. However, great advantage of reduction of Eq. (\ref{eqgz2}) to the form (\ref{eqg2b})
is in the fact that it allows to find the dependence of the function $g_{un}(z,\Delta)$ and of the eigenvalue $\eta_{un}^2$ on the mass ratio $\Delta$ analytically by Eqs.  (\ref{gtilde1}) and  (\ref{eta2un})  respectively. Therefore it considerably simplifies  the study the effect of the unequal masses. 
At $r=\frac{m_2}{m_1}\to\infty$  it allows to find the limiting value of the two-body contribution $N_2$ to the state vector.

\section{Binding energy}\label{be}
Presence of the mass-dependent factor $(1-\Delta^2)$ in the expression  (\ref{lamal}) for $\lambda$ through ${\cal C}$ 
drastically affects the spectrum relative to the equation (33) from \cite{Cutkosky} in which, by definition, $\lambda$ does not depend on masses.
As mentioned, this factor appears in Eq. (\ref{eqgz2}) and it does not appear in Eq. (\ref{eqg2b}) since it is cancelling as a result of transformations. 
The equation (33) from \cite{Cutkosky} differs from Eq. (\ref{eqg2b}) only by the constant at the front of the integral: instead of $\frac{\cal C}{\pi}$ it contains  
$\frac{\lambda}{(1-\Delta^2)}$. 
If $\Delta$ increases from 0 to 1 (the constituent mass $m_1$ increases up to infinity) and if it is assumed that $\lambda$ does not depend on masses, the "effective constant" $\frac{\lambda}{(1-\Delta^2)}$ increases up to infinity. Whereas, if it exceeds $2$, the eigenvalue $\eta_{eq}^2$
(i.e., $M^2$) for the ground state $n=1$  becomes negative.  Yet, this fact itself does not mean appearance of negative mass $M^2<0$ since $\eta_{eq}^2$,
in the non-equal masses problem, is an auxiliary quantity. The true eigenvalue is determined by $\eta_{un}^2$ via Eq. (\ref{eta2un}). However, for large enough $r$, $\eta_{un}^2$ also becomes negative.
The numerical calculations show that the negative second term $(1-\Delta^2)\eta_{eq}^2$
in Eq. (\ref{eta2un}) increases in the absolute value with increase of $r$ in spite of the factor  $(1-\Delta^2)$ which tends to zero. This leads, starting with some $r$, to negative value $\eta_{un}^2$. This unequal masses solution also corresponds to tachyon with $M^2<0$. 
For any small $\lambda$, there exists large $r$ providing the tachyon solution.
This large $r$ undoubtedly exists since for available ion masses $r>10^5$.
This situation - appearance of tachyonic solutions - looks unsatisfactory. 
This does not happen in the equation (\ref{eqg2b}), where ${\cal C}$ is fixed and if it is smaller than $2\pi$  \cite{Wick,Cutkosky}\footnote{In Refs. \cite{Wick,Cutkosky} it is shown that the condition $M^2>0$ (absence of tachyons) is fulfilled if the coupling constant $\lambda$, used in these papers,
satisfies the condition $\lambda<2$. For the equal masses case, described by  Eq. (\ref{eqg2b}), we have ${\cal C}=\pi\lambda$ (see  Eq. (\ref{lamal0}) from Appendix \ref{app1}, taken for $m_1=m_1$), that provides the condition 
${\cal C}<2\pi$.}. As mentioned, the dependence of $\eta^2_{un}$ on masses in Eq. (\ref{eqgz2})
 is completely determined by  $\Delta$ in Eq. (\ref{eta2un}).
By means of (\ref{Delta}), Eq. (\ref{eta2un}) is rewritten as
\begin{equation}\label{etaun}
\eta_{un}^2=1-\frac{4r}{(r+1)^2}(1-\eta_{eq}^2).
\end{equation}
We will increase $m_1$ (the ion mass), keep $m_2$ (the electron mass) fixed and see the variation of the binding energy.
Since $\eta_{un}=\frac{M}{m_1+m_2}$, $M=m_1+m_2-B_{un}$, we find the binding energy:
\begin{equation}\label{Ba}
\frac{B_{un}}{m_2}=(r+1)-\sqrt{(r-1)^2+4r\eta_{eq}^2},
\end{equation}
where $\eta_{eq}$ -- the eigenvalue of Eq. (\ref{eqg2b}) -- does not depend on masses. Therefore the formula (\ref{Ba}) determines analytically the dependence of the ratio $B_{un}/m_2$ on the ratio  $r=m_1/m_2$.

In the limit $m_1\gg m_2$ ($r\to\infty$) the formula (\ref{Ba}) gives $B_{un}=2m_2(1-\eta_{eq}^2)$, that is, with \mbox{$\eta_{eq}=(2m_2-B_{eq})/(2m_2)$},
\begin{equation}\label{Bun}
B_{un}=2\left(1-\frac{B_{eq}}{4m_2}\right)B_{eq},
\end{equation}
that for small binding energy $B_{eq}\ll m_2$ is reduced to 
\begin{equation}\label{Bun_inf}
B_{un}=2B_{eq}.
\end{equation}
That is, when mass of one of the particles increases up to infinity, the binding energy increases up to two times. But, as mentioned above, for fixed value of the Coulomb coupling constant ${\cal C}<2\pi$, there are no  tachyon solutions {\it for any mass ratio $r$}.

\section{Two-body Fock component}\label{N_2}
The BS amplitude is the matrix element taken from the T-product of two Heisenberg field operators (corresponding to electron and heavy ion in our case) between the state vectors corresponding to vacuum and to the 
bound system\footnote{We hesitate to call this bound system "atomic electron-ion system", since an atomic system is composed from the charge particles, whereas, as mentioned, the abnormal system is dominated by photons.}.
In the coordinate space it reads
\begin{equation} \label{bs2}
\Phi({\tt x_1,x_2},p)=\langle 0 \left| T(\varphi ({\tt x}_1)\varphi ({\tt x}_2))\right|p\rangle\ .
\end{equation}
The bound state vector $|p\rangle$ is represented by infinite number of the Fock sectors corresponding to the states with different numbers of particles. 
The coefficients of this decomposition are the two-, three-, $\ldots$ body wave functions - the Fock components. 
The Heisenberg operators $\varphi({\tt x}_{1,2})$ turn into the free ones on the quantization plane. Depending on the quantization scheme, this can be either the equal time plane $t=0$ (in the instant form quantization), or the light-front one (in the front form). We work in the explicitly covariant version of the light-front 
dynamics with the quantization plane  of general orientation defined by the equation $\omega\cd \tt x=0$, where $\omega=(\omega_0,\vec{\omega})$ with $\omega^2=0$.  The four-vector $\omega$ determines the orientation of the light-front plane in the Minkowski space.
For the product of two free operators $\varphi({\tt x}_{1})\varphi({\tt x}_{2})$, only the two-body Fock component of the state vector $|p\rangle$ contributes into the 
matrix element (\ref{bs2}). Therefore, the BS amplitude with the argument constrained to the light-front plane is related to the two-body Fock component
$\psi (k_1,k_2,p,\omega \tau)$ of the state vector defined on the light-front plane. The four-momenta -- the arguments of this wave function --  satisfy the conservation law $k_1+k_2=p+\omega\tau$
and are on the corresponding mass shells: $k_1^2=m_1^2$, $k_2^2=m_2^2$, $p^2=M^2$, $(\omega\tau)^2=0$, but off-energy shell. The value  
$\tau=((k_1+k_2)^2-M^2)/(2\omega\cd p)$ determines the measure of  deviation of the wave function from the energy shell
 (see for review \cite{cdkm}).

The traditional and convenient parametrization of the light front wave function uses the variables $\vec{k}_{\perp},x$, denoted below\footnote{The space-time coordinate is denoted ${\tt x}$ to not be confused with the momentum ratio $x$.}
as  $\vec{R}_{\perp}, x$. 
Construction of these variables in terms of the four-momenta $k_1,k_2,p,\omega \tau$ 
is explained in the Appendix \ref{app2}.
The  derivation of the relation between the light front wave function $\psi(\vec{R}_{\perp},x)$ and the BS amplitude $\Phi(k,p)$ in the momentum space is given in \cite{cdkm}. It has the form:
\begin{equation}\label{lfwf}
\psi(\vec{R}_{\perp},x)=\frac{x(1-x)}{\pi\sqrt{N^{un}_{tot}}} \int_{-\infty}^{\infty}\Phi\left(k+\frac{\beta\omega}{\omega\cd p},p\right) d\beta,
\end{equation}
where $N^{un}_{tot}=\langle p\vert p\rangle$. 

We will limit ourselves mainly to the ground state with $n=1$. Therefore, we substitute in Eq. (\ref{lfwf}) the BS amplitude (\ref{Phi1}). 
We get:
\begin{eqnarray}\label{wf1}
\psi(\vec{R}_{\perp},x)&=&-\frac{ix(1-x)}{\pi\sqrt{N^{un}_{tot}}}m_{12}^3
\int_{-1}^1g_{un}(z,\Delta)\,dz
\nonumber\\
&\times&\int_{-\infty}^{\infty} \frac{d\beta}{(\beta a+ b-i\epsilon)^3},
\end{eqnarray}
where $g_{un}(z,\Delta)$ is the solution of the unequal masses equations (\ref{eqgz2}), (\ref{equn}),
\begin{eqnarray*}
 a&=&1-2x-z,
\\
 b&=&\frac{1}{4x(1-x)}\{\vec{R}^2_{\perp}-x(1-x)M^2+m_{12}^2[(1+\Delta)^2
\\
&-&4x\Delta]\}[1-(1-2x)z-(1-2x-z)\Delta].
\end{eqnarray*}
To derive Eq. (\ref{wf1}), we need to express the scalar products $k_{1,2}\cd p$, $k_1\cd k_2$ which appear in (\ref{lfwf}) through the variables $\vec{R}_{\perp},x$.  These expressions are found in the Appendix \ref{app2}.

The integral over $\beta$ in (\ref{wf1}) gives the delta-function:
$$
\int_{-\infty}^{\infty} \frac{d\beta}{(\beta  a+ b-i\epsilon)^3}=\frac{i\pi}{ b^2}\delta( a).
$$
By means of this delta-function, we integrate over $z$.  Then $ b$ obtains the form:
$$
 b = \vec{R}^2_{\perp}+m_{12}^2Q(z,\Delta),
$$
where $z=1-2x$, $Q(z,\Delta)$ is defined in (\ref{Qt}).
In this way, we get:
\begin{equation}\label{wf2}
\psi(\vec{R}_{\perp},x)= 
\frac{(1-z^2)}{4\sqrt{N^{un}_{tot}}}\frac{m_{12}^3g_{un}(z,\Delta)}{[\vec{R}^2_{\perp}+m_{12}^2Q(z,\Delta)]^2},
\end{equation}
with
$N^{un}_{tot}=F(0)$. Appearance of the electromagnetic form factor  $F(0)$ instead of $N^{un}_{tot}$ and its calculation are explained
in Sec. \ref{norm}. $F(0)$ is given by Eq. (\ref{F2f}) below.

\section{Contribution $N_2$ of the electron-ion Fock sector to the total norm $N^{un}_{tot}$}\label{sect_N2}

The contribution $N_2$ of the two-body electron-ion Fock sector to the total norm $N^{un}_{tot}$ is determined by the two-body Fock component $\psi(\vec{R}_{\perp},x)$:
\begin{eqnarray}\label{N2}
N_2^{un}&=&\frac{1}{(2\pi)^3}\int|\psi^2(\vec{R}_{\perp},x)|^2\frac{d^2R_{\perp}dx}{2x(1-x)}
\nonumber\\
&=&\frac{1}{3\cd 2^7\pi^2N^{un}_{tot}}\int_{-1}^1\frac{(1-z^2)g_{un}^2(z,\Delta)dz}{Q^3(z,\Delta)},
\end{eqnarray}
We substituted in the  first integral of Eq. (\ref{N2}) the wave function (\ref{wf2}) and calculated the integral over $d^2R_{\perp}$.
The normalization of the function $g_{un}(z,\Delta)$ is not important since the normalization factor enters in Eq. (\ref{N2}) both in numerator and denominator (via 
$N^{un}_{tot}$) and cancels.

We replace in (\ref{N2}) the integration variable $z$ by $\bar{z}$ defined by Eqs. (\ref{zzbar}) 
and express $\Delta$ by Eq. (\ref{Delta}) via the ratio $r=m_1/m_2$.
In this way, Eq. (\ref{N2}) is transformed as:
\begin{eqnarray}\label{N2a}
N^{un}_2&=&\left.\frac{(1+r)^8}{3\cd 2^{15}\pi^2r^4N^{un}_{tot}}\int_{-1}^1\frac{(1-\bar{z}^2)g_{eq}^2(\bar{z})d\bar{z}}{[1-(1-\bar{z}^2)\eta_{eq}^2]^3}\right\vert_{r\to\infty}
\nonumber\\
&=&\frac{r^4}{3\cd 2^{15}\pi^2N^{un}_{tot}}\int_{-1}^1\frac{(1-\bar{z}^2)g_{eq}^2(\bar{z})d\bar{z}}{[1-(1-\bar{z}^2)\eta_{eq}^2]^3}.
\end{eqnarray}
The limit $r\to\infty$ is taken in order to get the leading degree of $r$.
We see that the dependence of $N^{un}_2$ on the mass ratio $r$ (without taking into account the $r$-dependence of the factor $N^{un}_{tot}$) is very simple and at large $r$ is reduced to the factor $r^4$. The asymptotical behavior of $N^{un}_{tot}$ at $r\to\infty$ will be found in the next section.

\section{Normalization of the BS amplitude}\label{norm}
Normalization of the state vector $\vert p\rangle$, entering in the definition (\ref{bs2}) of the BS amplitude, determines the normalization of this amplitude. 
In its turn, the latter determines the normalization of the electromagnetic form factor  of the whole system $F(q)$ ($q$ is the momentum transfer)which is expressed through the BS amplitude. $F(q)$ incorporates all the Fock components, but, as explained above, it is calculated taking into account the coupling of photon with electron only, but not with the ion.
The condition 
$\langle p\vert p\rangle=1$ is equivalent to $F(0)=1$ (see e.g. \cite{Nakanishi_prog}). Therefore, identifying, for arbitrary normalized state vector $\vert p\rangle$,  $N_{tot}=F(0)$ and dividing the state vector by $\sqrt{N_{tot}}$, we get the state vector and the electromagnetic form factor normalized to 1. This results in the factor $1/\sqrt{N_{tot}}$ in the expression (\ref{lfwf}) for $\psi(\vec{R}_{\perp},x)$.
 
To ensure the condition $F(0)=1$ with the state vector normalized to 1, the form factor should be calculated with true vector photon,  whereas the nature of forces providing the bound state is irrelevant. In the general case of different masses, the $q$-behavior of the electromagnetic form factor depends on the choice  which particle the photon couples to. However, at $q=0$ this dependence disappears, since in this case the photon does not probe the structure of system but gives information about its total charge only, which for a nucleus with the charge $Ze$ and electron is $(Z-1)e$. Thus, if we take the electron contribution only and remove the negative charge $-e$, we obtain the form factor $F(q)$ normalized to 1  at $q=0$. Therefore, it is convenient to consider only this contribution.
The electromagnetic form factor $F(q)$ is related to the electromagnetic  current $J_{\mu}$ of a bound system as $J_{\mu}=(p+p')_{\mu}F(q)$.
$J_{\mu}$  corresponds to the triangle graph which is calculated by the Feynman rules (see for details e.g. Refs. \cite{CKS21,cdkm}). 
Taking the contribution resulting from the interaction of the photon with the electron (the light particle No. 2), we obtain:
\begin{eqnarray}\label{F02}
F(0)=\frac{p\cd J}{2M^2}&=&\frac{i}{M^2}\int \frac{d^4k_1}{(2\pi)^4}(p\cd k_2)(k_1^2-m_1^2)
\nonumber\\
&\times&\bar{\Phi}\left(k,p\right)\Phi\left(k,p\right).
\end{eqnarray}
The factor $p\cd k_2$  results from the electromagnetic cur\-rent of the particle 2: $(k_{2\mu}+k'_{2\mu})$.

After expressing the momentum $k_{1}$ and $k_2$ via $k$ by Eq.  (\ref{k12}) the form factor  becomes:
\begin{eqnarray}\label{F02d}
F(0)&=&\frac{i}{M^2}\int \frac{d^4k}{(2\pi)^4}
(\mu_2 M^2-pk)
\nonumber\\
&\times&(\mu_1^2M^2+2\mu_1 pk+k^2-m_1^2)\bar{\Phi}\left(k,p\right)\Phi\left(k,p\right).
\nonumber\\
&&
\end{eqnarray}
The calculation of this integral is standard. Each BS amplitude in their product in Eq. (\ref{F02d}) is represented in the integral form (\ref{Phi1}), symbolically:
$$
\bar{\Phi}\left(k,p\right)\Phi\left(k,p\right)=\int_{-1}^1dz \ldots \frac{g(z)}{D^3(k,z)}\int_{-1}^1dz' \ldots \frac{g(z')}{D^3(k,z')},
$$
$D^3(k,z)$ is the denominator of the integrand in (\ref{Phi1}).
We use the relation:
$$
\frac{1}{D^3(k,z)D^3(k,z')}=\int_0^1\frac{30u^2(1-u)^2du}{[uD(k,z)+(1-u)D(k,z')]^6}.
$$
Then we shift the integration momentum $k=k'+cp$ and find $c$ from the condition that the terms linear in $k'$ are absent in the denominator 
$uD(k,z)+(1-u)D(k,z')$. By the Wick rotation $k'_0=ik'_4$ we transform the integral to the Euclidean space, where 
${k'}^2={k'}^2_0-\vec{k'}^2=-{k'}^2_4-\vec{k'}^2\to-{k'}^2_E$.
In the numerator, the odd degrees of the scalar products $k'\cd p$ can be omitted, the 2nd degree is replaced as $(k'\cd p)^2\to \frac{1}{4}M^2{k'}^2$ and the 4D integral (\ref{F02d}) is reduced to a 1D one: $d^4k'\to i2\pi^2{k'}_E^3dk'_E$.
Calculating the integral over $k'_E$ analytically, we find:
\begin{eqnarray}\label{F2f}
&&F(0)=\frac{m_{12}^6}{32\pi^2}\int_{-1}^1g_{un}(z',\Delta)dz' \int_{-1}^1g_{un}(z,\Delta)dz
\nonumber\\
&\times&
\int_0^1du\, u^2(1-u)^2
\nonumber\\
&\times&
\frac{m_1^2\xi(2+3\xi)-m_2^2(1-\xi)(1-3\xi)-2M^2\xi(1-\xi)}
{[m_1^2\xi+m_2^2(1-\xi)-M^2\xi(1-\xi)]^4},
\nonumber\\
&&
\end{eqnarray}
where 
$$
\xi=\frac{1}{2}(1+z)u+\frac{1}{2}(1+z')(1-u).
$$

For a test of our calculations, we also calculated  the form factor assuming that the photon interacts with the heavy particle 1.
The obtained formula has  a form which does not coincide,  at first glance, with Eq. (\ref{F2f}). However, the numerical value 
of $F(0)$ found in this way is the same as given by Eq. (\ref{F2f}).

We can find now the dependence of the form factor $F(0)$ on $r$ at $r\to\infty$ analytically.  To this aim, 
by Eq. (\ref{gtilde1}), we express in Eq. (\ref{F2f}) the function $g_{un}(z,\Delta)$ for non-zero $\Delta$, satisfying Eq. (\ref{eqgz2}),  via  the function $g_{eq}(z)$ for the equal masses  ($\Delta=0$), satisfying Eq. (\ref{eqg2b}). In addition, we represent the mass $M^2$ of the bound system for unequal constituent masses as
$M^2=4m_{12}^2\eta^2_{un}$  and express $\eta^2_{un}$  by means of Eq. (\ref{eta2un}) through the value 
$\eta^2_{eq}=\frac{M_{eq}^2}{4m^2}$ for equal masses $m_1=m_2=m$, for the same coupling constant ${\cal C}$. Since, after that, the integral contains the function $g_{eq}(z)$ which does not depend on $\Delta$, all the $\Delta$ dependence is hidden in the integrand, in the analytical factor multiplying the product $g_{eq}(z)g_{eq}(z')$. 
Decomposing this factor in series of $r$ at $r\to\infty$, we find
\begin{equation}\label{Ntot}
N^{un}_{tot}(r\to\infty)=\left.F(0)\right\vert_{r\to\infty}= Ar^4.
\end{equation}
The coefficient $A$ is given by a three-dimensional integral,  resulting from Eq. (\ref{F2f}). The
 degree $r^4$ is the same as in Eq. (\ref{N2a}) for $N_2$. Therefore, the factor $r^4$ in (\ref{N2a}) cancels and we find analytically that $N^{un}_2$ at $r\to\infty$ tends to a finite limit. This limiting value will be calculated numerically in the next section.

\section{Numerical results}\label{num}
We will study numerically the influence of the unequal masses, i.e., the dependence on the factor $r$ of the solution 
$g_{un}(z,\Delta)$ and corresponding  binding energy $B$, and of the two-body contribution $N_2$. All the calculations will be carried out for the coupling constant ${\cal C}=5$. The results for $B$ and $N_2$, for $n=1$ and $\kappa=0,2,4$, at $r=1,10,\infty$, are shown in the Table~\ref{tab1}. 
We see that if mass $m_1$ of the constituent 1 increases,  the binding energy also increases. In the limit $m_1\to\infty$ the binding energy increases 
twice relative to its value at $r=1$, in accordance with Eq.   (\ref{Bun_inf}).

\begin{figure}[h!]
\vspace{.5cm}
\begin{center}
\epsfxsize=8.2cm
\epsfbox{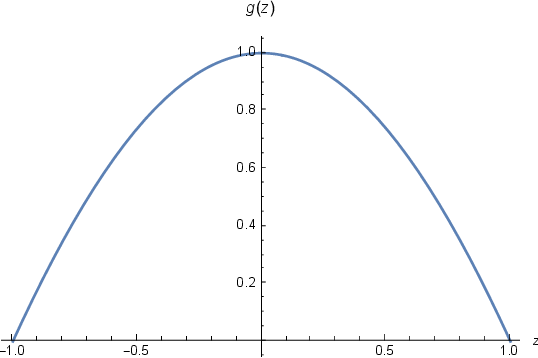}\\
\vspace*{0.5cm}
\epsfxsize=8.2cm
\epsfbox{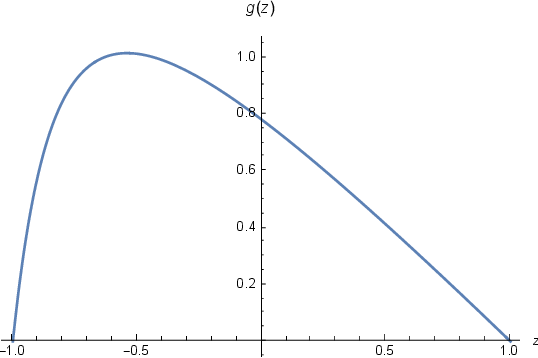}
\end{center}
\caption{Upper panel: $g_{eq}(z)$, equal masses ($r=1$)  for the normal state No. 1  of the Table~\ref{tab1}.\\
Lower panel: $g_{un}(z,\Delta)$, non-equal masses ($r=10\to\Delta=\frac{9}{11}$),  for the normal state No. 2  of the Table~\ref{tab1}.}
\label{fig1}
\end{figure}

\begin{figure}[h!]
\vspace{.5cm}
\begin{center}
\epsfxsize=8.2cm
\epsfbox{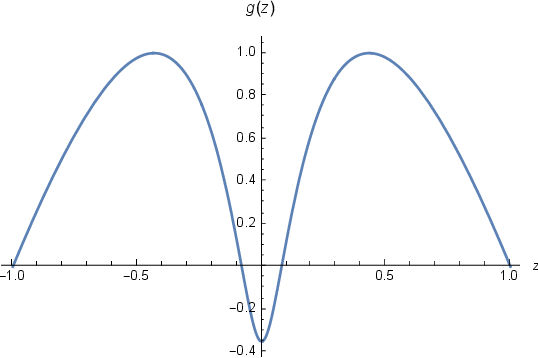}\\
\vspace{0.5cm}
\epsfxsize=8.2cm
\epsfbox{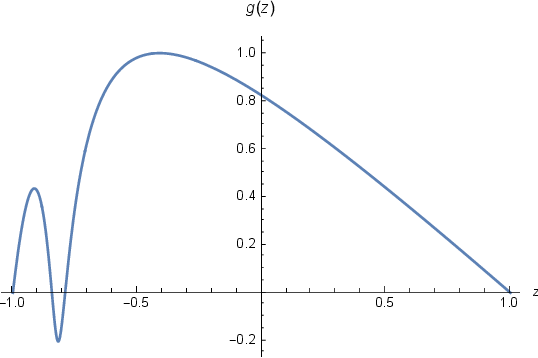}
\end{center}
\caption{Upper panel: $g_{eq}(z)$, equal masses ($r=1$)  for the abnormal state No. 4 ($\kappa=2$)  of the Table~\ref{tab1}.\\
Lower panel: $g_{un}(z,\Delta)$, non-equal masses ($r=10\to\Delta=\frac{9}{11}$),  for the abnormal state No. 5  ($\kappa=2$)  of the Table~\ref{tab1}.}
\label{fig2}
\end{figure}

\begin{figure}[h!]
\vspace{.5cm}
\begin{center}
\epsfxsize=8.2cm
\epsfbox{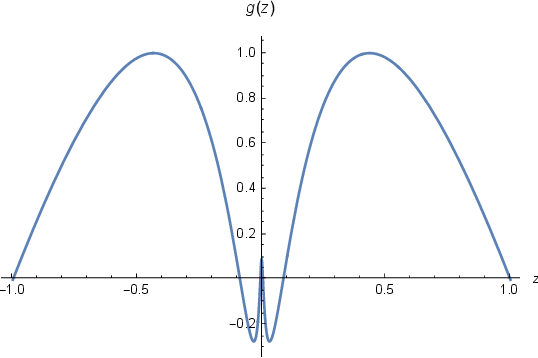}\\
\vspace{0.5cm}
\epsfxsize=8.2cm\epsfysize=5.cm
\epsfbox{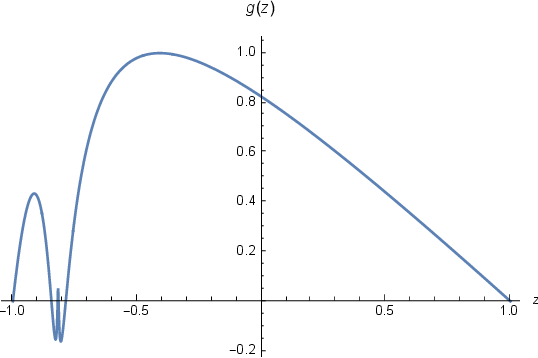}
\end{center}
\caption{Upper panel: $g_{eq}(z)$, equal masses ($r=1$)  for the abnormal state No. 7 ($\kappa=4$)  of the Table~\ref{tab1}.\\
Lower panel: $g_{un}(z,\Delta)$, non-equal masses ($r=10\to\Delta=\frac{9}{11}$),  for the abnormal state No. 8 ($\kappa=4$)   of the Table~\ref{tab1}.}
\label{fig3}
\end{figure}

\begin{figure}[h!]
\vspace{.5cm}
\begin{center}
\epsfxsize=8.2cm
\epsfbox{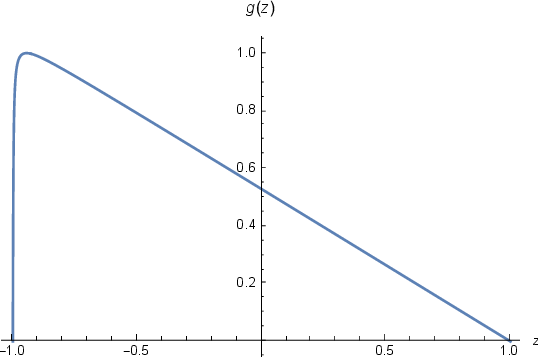}\\
\vspace{0.5cm}
\epsfxsize=8.2cm
\epsfbox{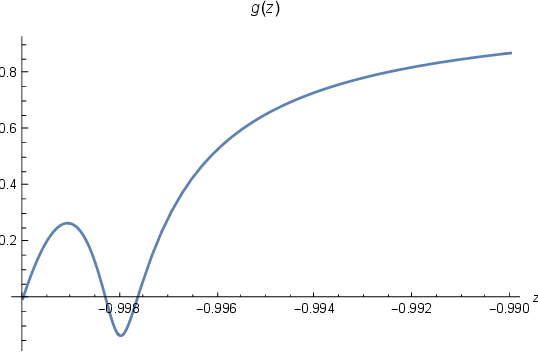}\\
\vspace{1cm}
\epsfxsize=8.2cm\epsfysize=5.cm
\epsfbox{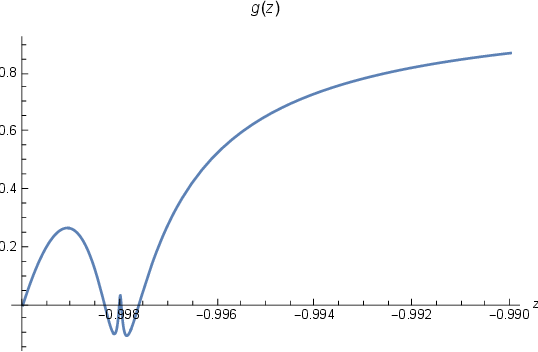}
\end{center}
\caption{$g_{un}(z,\Delta)$ for $r=10^3$.  
\\
Upper panel: for $\kappa=0,2,4$ (the curves are indistinguishable from each other in this scale).\\
Middle panel: $\kappa=2$, in the interval $-1\leq z\leq -0.99$. \\
Lower panel: $\kappa=4$, in the interval $-1\leq z\leq -0.99$.}
\label{fig4}
\end{figure}

The solutions of the equations (\ref{eqg2b}), (\ref{eqeq})  for $g_{eq}(z)$ and the equations (\ref{eqgz2}), (\ref{equn}) for $g_{un}(z,\Delta)$ at $n=1,\;\kappa=0,2,4$ and for $r=1,10,10^3$
are shown in Figs. \ref{fig1}-\ref{fig4}. All these solutions are normalized so that their maximal values achieved in the figures are equal to 1. The numerically found binding energies 
and the solutions $g_{un}(z,\Delta)$  exactly coincide with those found from the relations (\ref{eta2un}) and (\ref{gtilde1}). The number of nodes of the function 
 $g_{un}(z,\Delta)$ vs. $z$ is the same as for  $g_{eq}(z)$ and still coincides with the quantum number $\kappa$ though they are shifted to the smaller values of $z$.

\begin{table}[h!]
\begin{center}
\caption{The ratios $B/m_2$  and two-body norms $N_2$
for the principal quantum number $n=1$, for 
the normal ($\kappa=0$) and abnormal
($\kappa=2$, $\kappa=4$) states, for the coupling
constant value ${\cal C}=5$ and for equal ($r=1$) and unequal ($r=10,\infty$) masses.}\label{tab1}
\begin{tabular}{lllll}
\hline\noalign{\smallskip} 
No.&    $\kappa$ & r &  $B/m_2$ & $N_2$  \\
\noalign{\smallskip}\hline 
\noalign{\smallskip}
1&0  & 1 &   0.99926         &0.65\\
2&0  & 10&  1.45983         & 0.65\\
3&0  & $\infty$&  1.99852         & 0.65\\
\noalign{\smallskip}\hline
\noalign{\smallskip}
4& 2  & 1 & $3.51169 \cdot 10^{-3}$& 0.094 \\
5& 2 & 10 & $6.38114 \cdot 10^{-3}$& 0.102 \\  
6  & 2 & $\infty$ &$7.02338 \cdot 10^{-3}$ &0.093\\
\noalign{\smallskip}\hline
\noalign{\smallskip}
7&4& 1 &$1.54091\cdot 10^{-5}$ & $6.19\cdot 10^{-3}$\\
8&4& 10 &$2.80165    \cdot 10^{-5}$ &$6.86\cdot 10^{-3}$\\
9  &4& $\infty$& $3.08182   \cdot 10^{-5}$ &$6.67\cdot 10^{-3}$\\ 
\noalign{\smallskip}\hline
\end{tabular}
\end{center}
\end{table}



\section{Conclusions}\label{concl}
The spectrum of the system of two charged particles with  large  enough product of charges ${\cal C}>\frac{\pi}{4}$, in addition to the Coulomb levels, contains also infinite series of highly excited levels called abnormal \cite{Wick,Cutkosky}. These states are 
 dominated 
by Fock states containing photons in addition to the two massive charged particles, whereas the contribution of Fock state for two particles only is small~\cite{CKS21}. Therefore  these many-photons states are not  predicted by the Schr\"odinger equation.
They arise due to transformation, in the strong Coulomb field,  of an atomic state of two charged particles, interacting by exchange by single photons,  
to the states dominated by photons.
 The theoretical prediction and understanding the nature of the abnormal states put the experimental detection of such states on the agenda. To detect the abnormal states,
it would be ideal to carry out experiments with nucleus-antinucleus ions. In this case, the ions and antiions with $Z=11$ (Natrium and Antinatrium) would be enough to create a system with
the product of charges ${\cal C}=\frac{1}{137}Z^2\approx 0.88>\frac{\pi}{4}\approx 0.79$. In absence  at our disposal of antinuclei with  large enough charge, the experiments should be carried out with heavy ions and electrons. Then to provide ${\cal C}=\frac{1}{137}Z>\frac{\pi}{4}$ we need the ions with $Z\geq 108$ (Hassium and subsequent elements). 
Heavy ion and electron have very different masses. That's why we need to extend the  searchers \cite{Wick,Cutkosky,CKS21} for the case of constituents with unequal masses.

It turned out that the effect of unequal masses is "attractive" -- the binding energy increases when the mass ratio $r=\frac{m_1}{m_2}$ increases.
In the limit $r\to\infty$ (which, in practice, is very close to the mass ratio of heavy ion and electron), the binding energy increases up to  a factor two. The two-body constituent contribution $N_2$ is changing insignificantly and remains small. That is, the abnormal states are still dominated by the massless exchanges. They manifest themselves not only in deviation from the Coulomb spectrum, but also in special behavior of the electromagnetic form factors \cite{CKS21}. 

The same effect -- increase of the Coulomb binding energy by a factor of two when mass of one particles increases up to infinity -- takes place in the non-relativistic Schr\"odinger equation, where the binding energy is proportional to the reduced mass  $m^*=\frac{m_1 m_2}{m_1+m_2}=\frac{r m_2}{r+1}$. For $m_1=m_2$ ($r=1$): $m^*=\frac{1}{2}m_2$, whereas at $r\to \infty$: $m^*\to m_2$ - is two times larger. This 
cannot be expected in advance since the dynamics of two systems - abnormal and the Coulomb non-relativistic ones, governed by the corresponding equations,  are very different and the relation (\ref{Ba}), in general, does not provide the dependence on the reduced mass. 

Let us note that there exists the maximal value of the ion's electric charge (the critical charge $Z_c$). If  the ion charge exceeds this value, then an electron-positron pair is created in  the strong electric field. The electron is absorbed by the large ion charge,  not allowing it to exceed the critical value, while the  positron goes to infinity. 
This critical value is $Z_c=137$ for the point-like charge and it increases up to $Z_c=170$ due to finite size of nucleus (see for review \cite{ZP71}). The ion charge $Z>137\pi/4\approx 108$, needed for existence of the abnormal states in the system ion-electron, is less than the critical value $Z_c$ and therefore it is not forbidden. The interval $108\leq z\leq 170$ is rather wide, though the number of known long-living  enough isotopes suitable for experiments is restriced.

We did not take into account spins of constituents and photons. Spin-spin interaction can change the positions of the levels (the fine structure is not "fine" in the strong field), but it does not weaken the electric field. Moreover, the particles with opposite spins (e.g., electron and positron) interact stronger than the spinless ones.  Therefore, there is no reason to expect that the spins can eliminate the abnormal states. 

The interaction creating the abnormal states is not necessary electric. This could be exchange by gluons between quarks. In this case, the strong coupling constant can be enough for creation of the abnormal states which could have the hybrid nature \cite{CKS21}. 

The abnormal solutions for massive particle exchange (though for small exchange masses) also exist. Our study of them is in progress. 

In any case, the predicted abnormal states deserve the experimental study.

\appendix

\section{Relation between $\lambda$ and ${\cal C}$}\label{app1}

For any elastic amplitude $K$ calculated by the Feynman rules the cross section (in the c.m. frame) reads
$$
\frac{d\sigma}{d\Omega}=\frac{1}{64\pi^2s}|K|^2=\frac{1}{64\pi^2(m_1+m_2)^2}|K|^2.
$$
We follow the conventions given in the book \cite{IZ}. For simplicity we consider the low energy system since the relation between the coupling constants does not depend on the energy. Substituting here $K$ defined by Eq. (\ref{kern0}) we find:
\begin{equation}\label{sig1}
\frac{d\sigma}{d\Omega}=\frac{\pi^2\lambda^2(m_1+m_2)^2}{4q^4}.
\end{equation}

On the other hand, by definition,  ${\cal C}$ is the constant determining the Coulomb
potential:
$$
V(r)=-\frac{{\cal C}}{r}.
$$

The well known Born amplitude reads:
$$
f=\frac{m^*}{2\pi}V(\vec{q}),
$$
where $m^*=\frac{m_1m_2}{m_1+m_2}$ and
$$
V(\vec{q})=\int_0^{\infty}V(r)\exp(-i\vec{q}\cd\vec{r})d^3r=
\frac{4\pi{\cal C}}{q^2}.
$$
It is related to the cross section as:
$$
\frac{d\sigma}{d\Omega}=|f|^2.
$$

Hence 
\begin{equation}\label{sig2}
\frac{d\sigma}{d\Omega}=\frac{4m_1^2m_2^2{\cal C}^2}{(m_1+m_2)^2 q^4}.
\end{equation}
Comparing the cross sections (\ref{sig1}) and (\ref{sig2}), we find:
\begin{equation}\label{lamal0}
\lambda=\frac{4m_1 m_2}{\pi(m_1+m_2)^2}{\cal C}
\end{equation}
that can be rewritten in the form (\ref{lamal}). For equal masses and ${\cal C}=\alpha$ it is reduced to the relation $\lambda=\frac{\alpha}{\pi}$ used in \cite{CKS21}.

\section{Variables $\vec{R}_{\perp},x$ and the scalar products}\label{app2}
The light-front variables $\vec{R}_{\perp},x$  are constructed as follows.
We define the four-vectors $R_1=k_1-x_1p$, $R_2=k_2-x_2p$ with $x_{1,2}=(\omega\cd k_{1,2})/(\omega\cd p)$, $x_1+x_2=1$ and chose their components as: $R_{1}=(R_{10},\vec{R}_{1\parallel},\vec{R}_{1\perp})$ with $\vec{R}_{1\parallel}\parallel\vec{\omega}$, $\vec{R}_{1\perp}\cd\vec{\omega}=0$.
And similarly for $R_2$.
Since $\omega\cd R_1=\omega_0(R_{10}-R_{1\parallel})=0$, we get $R_{10}=R_{1\parallel}$, therefore $R_1^2=-\vec{R}_{1\perp}^2$.
The relation $R_1+R_2=\omega\tau$ gives $\vec{R}_{1\perp}=-\vec{R}_{2\perp}$ and $R_1\cd R_2=\vec{R}_{\perp}^2$. 
We denote $\vec{R}_{1\perp}\equiv \vec{R}_{\perp}$, $x_1\equiv x$. Then the light front wave function is parametrized as $\psi=\psi(\vec{R}_{\perp},x)$.

To express the scalar product $k_{1}\cd p$  through the variables $\vec{R}_{\perp},x$, we take square:
$R_1^2=-\vec{R}_{\perp}^2=m_1^2-2x(k_1\cd p)+x_1^2M^2$. From here we find the scalar product $k_{1}\cd p$ (the four-vectors $k_1,k_2$ are on the corresponding mass shells $k_{1,2}^2=m_{1,2}^2$). 
Similarly for $k_2\cd p$. To find  
$k_1\cd k_2$, we take the product $\vec{R}_{\perp}^2=R_1\cd R_2=(k_1-x_1p)\cd(k_2-x_2 p)$. In this way, we find the following expressions for the scalar products:
\begin{eqnarray*}
k_1\cd p&=&\frac{1}{2x_1}(\vec{R}^2_{\perp}+m_1^2+x_1^2M^2),
\\
k_2\cd p&=&\frac{1}{2x_2}(\vec{R}^2_{\perp}+m_2^2+x_2^2M^2),
\\
k_1\cd k_2&=&\vec{R}^2+x_1(k_2\cd p)+x_2(k_1\cd p)-x_1x_2M^2,
\\
k^2&=&\mu_2^2 m_1^2-2\mu_1\mu_2 (k_1\cd k_2)+\mu_1^2 m_2^2,
\end{eqnarray*}
where $k=\mu_2 k_1-\mu_1 k_2$ and $\mu_{1,2}$ are defined in (\ref{k}).
Using these relations, we can find the scalar products appearing in  (\ref{lfwf}):
\begin{eqnarray*}
\left(k+\frac{\beta\omega}{\omega\cd p}\right)^2&=&k^2+2(\mu_2 x_1-\mu_1x_2)\beta,
\\
p\cdot \left( k+\frac{\beta\omega}{\omega\cd p}\right)&=&\mu_2k_1\cd p-\mu_1k_2\cd p+\beta.
\end{eqnarray*}



\begin{thebibliography}{10}

\bibitem{bs}   E.E.~Salpeter,  H.~Bethe, A Relativistic Equation for Bound-State Problems.  Phys. Rev.  {\bf 84},  1232 (1951)

\bibitem{Wick}G.C. Wick, 
Properties of the  Bethe-Salpeter Wave Functions.
Phys. Rev. {\bf 96}, 1124 (1954)

\bibitem{Cutkosky}
R.E.~Cutkosky,  Solutions of the Bethe-Salpeter Equation.
Phys. Rev. {\bf 96},  1135 (1954)

\bibitem{Nakanishi_prog}H.~Nakanishi, A General Survey of the Theory of the Bethe-Salpeter Equation, Prog. Theor. Phys. Suppl., {\bf 43}, 1 (1969)

\bibitem{CKS21}J.~Carbonell, V.A.~Karmanov, H.~Sazdjian, 
Hybrid nature of  the abnormal solutions of the Bethe-Salpeter
equation in the Wick-Cutkosky model. Eur. Phys. J. C {\bf 81}, 50 (2021)

\bibitem{cia}
M.~Ciafaloni, P.~Menotti, Operator analysis of the
Bethe-Salpeter equation. Phys. Rev. {\bf 140}, B929 (1965)

\bibitem{cdkm}
J.~Carbonell, B.~Desplanques, V.A.~Karmanov and \mbox{J.-F.}~Mathiot,
Explicitly Covariant Light-Front Dynamics and Relativistic Few-Body Systems.
Phys. Reports, {\bf 300},  215 (1998)

\bibitem{ZP71} Ya.B. Zeldovich, V.S. Popov, Electronic structure of superheavy atoms. Uspekhi Fiz. Nauk, {\bf 105}.  403 (1971) [transl.: Sov. Phys. Usp. 
{\bf 14}, 673 (1972)] 

\bibitem{IZ} C.~Itzykson and J.-B.~Zuber,  {\it Quantum Field Theory} (Dover Publications, New York, 1980)


 \end{thebibliography}
 \end{document}